\documentclass{article}    % Specifies the document style.

\usepackage{graphicx}% Include figure files
\usepackage{dcolumn}% Align table columns on decimal point
\usepackage{bm}% bold math

\title{Is the pioneer anomaly a counter example to the dark matter hypothesis?}
\author{\textbf{Firmin J. Oliveira}\\
 Joint Astronomy Centre,\\
 Hilo, Hawai`i, USA 96720\\
\textit{firmin@jach.hawaii.edu}}

\begin{document}
\maketitle  

\begin{abstract}
  The Hubble law is extended to  massive particles based on the de Broglie
  wavelength. Due to the expansion of the universe the wavelength of an
  unbound particle would increase according to its cosmological redshift.
  Based on the navigation anomalies of the Pioneer 10 \& 11 spacecraft it
  is postulated that an unbound massive particle has a cosmological
  redshift $z = (c / v_0)\, H_0\, t$, where $c$ is the speed of light
  in vacuum, $v_0$ is the initial velocity of the particle, $H_0$ is
  Hubble's constant and $t$ is the duration of time that the particle
  has been unbound. The increase in wavelength of the particle
  corresponds to a decrease in its speed by $\Delta{v} = - c\, H_0 \, t$.
  Furthermore, it is hypothesized that the solar system has escaped the
  gravity of the Galaxy as evidenced by its orbital speed and radial
  distance and by the visible mass within the solar system  radius.
  This means that spacecraft which become unbound to the solar system
  would also be galactically unbound and subject to the Hubble law.
  This hypothesis and the extended Hubble law may explain the anomalous
  acceleration found to be acting upon the unbound Pioneer  10 \& 11
  spacecraft. Thus, the Pioneer anomaly may be a counter example to the
  dark matter hypothesis.

  Because photons have a speed which make them unbound to the Galaxy,
  it is predicted that the navigation beam in open space would undergo
  a cosmological redshift in its frequency which would be detectable
  with modern clocks.

\end{abstract}


\section{\label{sec:Intro} Introduction}

 In this paper we will attempt to explain the anomalous acceleration found in the
 Pioneer 10 \& 11 spacecraft as being due to the expansion of the universe. To
 accomplish this task, the Hubble law, which applies to the wavelength of light
 from distant galaxies, is extended to include unbound massive particles.  It is
 well known that microscopic particles by way of their de Broglie wavelength
 display wave interference phenomena identical to light waves. We argue that the
 expanding universe, by the Hubble law, can have an effect on the de Broglie
 wavelength of a galactically unbound massive particle analogous to the way it
 increases the wavelength of a photon. The increase in wavelength of the particle
 corresponds to a decrease in its velocity.

 In order for the Pioneer spacecraft to be unbound to the Galaxy it is necessary
 to hypothesize that the solar system itself is already galactically unbound.
 Evidence is given in support of this concept.

\section{Newtonian system in an expanding universe}

 In a paper by Anderson\cite{JLAnderson1995} it was found that
 a Newtonian system of neutrally or electrically charged particles
 in an expanding Einstein-deSitter universe can be described by
 (Ref. \cite{JLAnderson1995}, Eq. 9)
 \begin{eqnarray}
  m_A \partial^2_{t_S} {\bf x}_A
      \,\,\, + \,\,\, 2 \frac{\epsilon_H}{\epsilon_S} \partial_{t_S} \partial_{t_H} m_A {\bf x} &=&
      - \frac{1}{R^3} \sum_{B \ne A} {\frac{m_A m_B}{r^3_{AB}} {\bf r}_{AB}} \nonumber \\
  & & - 2 \frac{\epsilon_H}{\epsilon_S} m_A \frac{\partial_{t_H} R}{R} \partial_{t_H}{\bf x}_A
      \nonumber \\
  & & + {\cal{O}}(\epsilon^2_S,\epsilon^2_H), \label{eq:JLAnderson-1}
 \end{eqnarray}
 where there are particles labeled $A, B, C, \ldots,$ in the multiparticle system,
 $m_A$ is the mass of the $A^{th}$ particle whose coordinates are the vector ${\bf x}_A$,
 $m_B$ is the mass of the $B^{th}$ particle whose coordinates are the vector ${\bf x}_B$,
 etc., and ${\bf r}_{AB} = {\bf x}_A - {\bf x}_B$.  We have also the scale factor
 $R(t_H)=(t_H/t_0)^{2/3}$ with $t_H$ the parametrized cosmic time, $t_0$ the time at the
 present epoch where the Hubble time at the current epoch is
 $T_H = R(t)/ {\dot{R}}(t) = (3/2) t_0$.  The short-hand form of the partial derivative
 is expressed $\partial_{t} = \partial / \partial{t}$.  
 Defining the time $T_L$ as the typical time for light to travel across the system
 of particles, and the time $T_S$ as a characteristic time (typical orbital period) of
 the system, then $\epsilon_S = T_L / T_S$ and $\epsilon_H = T_L / T_H$, and from these
 are defined the\ times $t_S = \epsilon_S t$ and $t_H = \epsilon_H t$ in terms of the
 cosmic time $t$.

 Anderson made two conclusions from (\ref{eq:JLAnderson-1}).  First, if the system of 
 particles is small such that $\epsilon_H \ll \epsilon_S$ then 
 \begin{equation}
   r^3 \omega \,\, = \,\, \frac{\alpha}{R^3} \,\,\,\,
       {\rm and} \,\,\,\, r \, \omega \,\, = \,\, \frac{\beta}{R},
      \label{eq:Anderson-2}
 \end{equation}
 where $\alpha$ and $\beta$ are constants, $r$ is the comoving coordinate distance between
 the masses, and $\omega$ is the orbital angular velocity. From this it follows that
 \begin{equation}
  \omega \,\, = \,\, {\rm const} \,\,\,\,
     {\rm and} \,\,\,\, R  r \,\, = \,\, {\rm const} ,
      \label{eq:Anderson-3}
 \end{equation}
 which means that the orbital frequencies $\omega$ and radial distances $ R r$ of the
 particles in the system are not changed by the expanding universe.  This would be the
 situation for a small bound system of particles where Newton's laws apply, which is
 obtained from (\ref{eq:JLAnderson-1}) with $\epsilon_H \ll \epsilon_S$,
 \begin{equation}
     m_A \partial^2_{t_S} {\bf x}_A \,\, = \,\,
     - \frac{1}{R^3} \sum_{B \ne A} {\frac{m_A m_B}{r^3_{AB}} {\bf r}_{AB}}.
        \label{eq:Newtons_law}
 \end{equation}

 In the other case when $\epsilon_H \approx \epsilon_S$ there is an effect of the
 expansion upon the system of particles due to the two terms coupled by
 $\epsilon_H / \epsilon_S$ in (\ref{eq:JLAnderson-1}).
 For example, this conditon is satisfied by an unbound two particle system of a large
 mass (Sun) and a small mass (comet or spacecraft) which is in hyperbolic orbit, because
 in this case the system size is ``infinite'' so that the light travel time $T_L$  across
 the system and the characteristic period $T_S$ of the system  are both comparable to
  the Hubble time $T_H$. Then the second term on the right hand side of
 (\ref{eq:JLAnderson-1}) works out to
 \begin{equation}
 - 2 \frac{\epsilon_H}{\epsilon_S} m_A \frac{\partial_{t_H} R}{R}
        \partial_{t_H} m_A {\bf x}_A \,\, = \,\, 
      - \frac{4}{3}\frac{\epsilon_H}{\epsilon_S}
           m_A \frac{\partial_{t_H}{\bf x}_A}{t_H} ,
    \label{eq:2nd-term-rhs}
 \end{equation}
 where
 \begin{equation}
    \frac{\partial_{t_H} R}{R} \,\, = \,\,
      \frac{\partial_{t_H} \left(t_H/t_0\right)^{2/3}} {\left(t_H/t_0\right)^{2/3}} 
      \,\, = \,\, \frac{2}{3} \frac{1}{t_H} .
 \end{equation}
 Substituting this result into (\ref{eq:JLAnderson-1}), and assuming that
 $\partial_{t_S} \partial_{t_H} m_A {\bf x} = 0$, the equation of motion of a
 particle of mass $m_A$ in unbound orbit (when $\epsilon_H \approx \epsilon_S$) is
 given by
 \begin{eqnarray}
    m_A \partial^2_{t_S} {\bf x}_A \,\, = \,\,
          -  \frac{1}{R^3} \sum_{B \ne A} {\frac{m_A m_B}{r^3_{AB}} {\bf r}_{AB}}
         \,\, - \,\, \frac{4}{3}\frac{\epsilon_H}{\epsilon_S}
               m_A \frac{\partial_{t_H}{\bf x}_A}{t_H} . \label{eq:pioneer_eq_2}
 \end{eqnarray}

 In the next section we look at what the Hubble law says about massive particles,
 and then make a connection with (\ref{eq:pioneer_eq_2}). 

\section{\label{sec:HubbleLaw} Hubble law for a massive particle}

The de Broglie\cite{deBroglie-1} wavelength $\lambda$ for a massive particle
 of momentum $p$ is given by
\begin{equation}
  \lambda \,\, = \,\, \frac{h}{p} , \label{eq:lambda0_massive}
\end{equation}
where $h$ is Planck's constant. This relation is also true for photons.

 In the expanding universe we assume that the Hubble law for photons is also valid for
 galactically unbound massive particles. Then, for a particle which had an initial wavelength
 of $\lambda_0$ at (cosmological) redshift 0 but is now at redshift $z$, its wavelength $\lambda$
 is given by
\begin{equation}
   \lambda \,\, = \,\, \left( 1 + z \right) \lambda_0 . \label{eq:lambda_massive}
\end{equation}
By (\ref{eq:lambda0_massive}) this can be written in momentum form
\begin{eqnarray}
   \frac{h}{p} \,\, &=& \,\, \left( 1 + z \right) \frac{h}{p_0} ,  \label{eq:h/p} \\
  \nonumber \\
   p \,\, &=& \,\, \frac{p_0}{ 1 + z } , \label{eq:p} \\
  \nonumber \\
  \Delta{p} \,\, &=&  \,\, p - p_0 \,\, = \,\, \frac{-z\, p_0}{1 + z} . \label{eq:delta_p_precise}
\end{eqnarray}
For $z \ll 1$ this implies
\begin{equation}
  \Delta{p} \,\, \approx \,\, -z\, p_0  . \label{eq:Delta_p_approx}
\end{equation}

The conclusion of the Pioneer 10 \& 11 report (Ref. \cite{Anderson2002}, Eq. 54)
 was that there was an anomalous
constant acceleration on the spacecraft directed toward the Sun of magitude
 $(8.74 \pm 1.33) \times 10^{-8} {\rm cm} \,{\rm s}^{-2}$. It was also noted
 (Ref. \cite{Anderson2002}, Sect. C) that this value is approximately equal to
 the speed of light times the Hubble constant, $c\, H_0$. During these measurements
 the motion of either spacecraft was directed nearly radially outward from the Sun,
 hence the reported direction of the anomalous acceleration is also consistent
 with it being directed along the line of and in opposition to the motion of the
 spacecraft. With these experimental facts, we postulate that:

 (P-1) {\em For $z \ll 1$ the change in momentum of an unbound particle due to the
 expansion of the universe is given by}
\begin{equation}
 \Delta{p} \,\, \approx  \,\, -z\, p_0 \,\, = \,\, - m_0 \, c \, H_0 \, t  ,
  \label{eq:Delta_p_Law}
\end{equation}
{\em where $m_0$ is the mass of the particle at redshift 0 and $t$ is the duration of time
 since the particle has been unbound.}

This implies that for a massive particle the redshift
\begin{eqnarray}
   z     \,\, &=& \,\, \frac{m_0 \, c}{p_0} \, H_0 \, t .  \label{eq:z_massive}
\end{eqnarray}
For a photon the momentum is $p_0=m_0 \,c$, so (\ref{eq:z_massive}) gives 
\begin{equation}
   z \,\, =  \,\, H_0 \, t \,\, = \,\, \frac{H_0}{c} \, r \, , \label{eq:photon_Law}
\end{equation}
which is recognized as the Hubble law for the redshift of light observed from a galaxy
at distance $r=c\,t$, where $t$ is the duration of time since the (unbound) photon
left the galaxy.

For a massive particle moving at non-relativistic speed,
 $m \approx m_0$, so divide (\ref{eq:Delta_p_Law}) by the mass to get the
cosmological velocity shift
\begin{equation}
  v_{z}(t) \,\, = \,\, \Delta{v} \,\, = \,\, -z\, v_0  \,\,\, = \,\,\, - c\, H_0 \, t , \label{eq:v_z} \\
\end{equation}
from which we get the redshift relation for a massive particle in terms of its
 initial velocity
\begin{equation}
      z \,\, = \,\, \frac{c}{v_0} \, H_0 \, t .  \label{eq:massive_particle_Law}
\end{equation}
For larger redshift, though still assuming non-relativistic velocities,
 from (\ref{eq:delta_p_precise}) and (\ref{eq:massive_particle_Law}) we {\em assume} that 
the cosmological velocity shift is given by
\begin{equation}
   v_{z}(t) \,\, = \,\, \frac{-z\, v_0}{1 + z}
    \,\, = \,\, \frac{-c \, H_0\, t }{ 1 + \left( c / v_0\right) H_0 \, t }. \label{eq:delta_v_precise}
\end{equation}
 Assuming that the Hubble law for massive particles is related to (\ref{eq:pioneer_eq_2})
 then from (\ref{eq:v_z}) we will assume the relation
 \begin{equation}
   \frac{v_{z}(t)}{t} \,\,\, = \,\,\, - c\, H_0 =
        \,\,\, - \frac{4}{3}\frac{\epsilon_H}{\epsilon_S}
           \frac{\mid \partial_{t_H}{{\bf x}_A \mid}}{t_H}.  \label{eq:a_P_equivalence} 
 \end{equation}
 Inherent in the assumption of (\ref{eq:a_P_equivalence}) is that the rate of change of
 the position vector $\partial_{t_H}{{\bf x}_A}$ is decreasing, hence the negative sign
 multiplies the magnitude of the vector.

 The final thing we need in our theory is a reason why a spacecraft which is unbound
 to the solar system would also be unbound to the Galaxy and therefore be in the category
 $\epsilon_H \approx \epsilon_S$ and governed by (\ref{eq:pioneer_eq_2}).
 
\section{Hypothesis that the solar system has escaped the gravity of the Galaxy} 

The solar system is at a distance\cite{Eisenhauer2003} of $R_0 = (8 \pm 0.4) \,{\rm Kpc}$
 from the center of the Galaxy.
 Its orbital circular speed is $V_0 = (220 \pm 15) \,{\rm km/s}$.
 The total visible mass $M_{vis}$ of the Galaxy (Ref. \cite{Dehnen-1}, Table 4, Model 4)
 is composed of the Galaxy's bulge and disk masses,
 with $M_{bulge} \approx 0.364 \times 10^{10} \, M_{\odot}$ and
 $M_{disk} \approx 4.16 \times 10^{10} \, M_{\odot}$, giving
\begin{equation}
M_{vis} \,\, \approx \,\, (4.52 \pm 0.90) \times 10^{10}  M_{\odot} , \label{eq:M_visible}
\end{equation}
where the error in (\ref{eq:M_visible}) is this author's expression of an uncertainty
of $\pm 20\, \%$. The total visible mass within the Sun's orbit would be less than this.
 If we assume that $V_0$ is greater than the escape velocity
 at $R_0$ then this implies that the Galaxy mass $ M_0$ within the
 solar radius would be
\begin{equation}
   M_0 \,\, < \,\, \frac{V^2_0 \, R_0}{2 G}
   \,\, = \,\, (4.55 \pm 0.84 ) \times 10^{10} \, M_{\odot} . \label{eq:M(Rsolar)}
\end{equation}
The right hand side approximately equals the total visible mass in (\ref{eq:M_visible}),
suggesting that the solar system velocity $V_{0}$ is greater than the escape velocity
for the entire visible Galaxy.

This discussion suggests that it is not unreasonable to hypothesize that:

 \noindent (H-1) {\em The solar system Galactic orbital speed exceeds escape velocity at its
  current radius.}

 This hypothesis avoids the need for any substantial halo of dark matter to
 bind the solar system with velocity $V_{0}$. This suggests that the Sun is in the Galaxy
 because, according to (\ref{eq:delta_v_precise}), the expansion of the universe over time
 continually decreases the solar system's velocity. This may help to account for why a galaxy
 does not fly apart even though the orbital speeds of objects in it exceed the Newtonian escape
 velocity of the visible mass. However, the theory of the dynamics of spiral galaxies is beyond
 the scope of this paper (cf., \cite{milgrom1988} and \cite{hartnett2006}.)

\section{Pioneer 10 \& 11 anomalous velocities and accelerations}

 By the hypothesis (H-1) the solar system is unbound to the Galaxy.
 This implies that any spacecraft which escapes from the solar system
 will also have escaped from the Galaxy. The Pioneer 10 \& 11 spacecraft\cite{Anderson2002}
 have escaped the solar system gravity and therefore by (H-1) are unbound to the
 Galaxy. Hence, their velocities will be cosmologically shifted as in
 (\ref{eq:v_z}), so that their anomalous velocity and acceleration are respectively
 given by
 \begin{eqnarray}
  v_{P}(t) \,\, &=& \,\,  v_z(t) \,\, = \,\, -c \, H_0 \, t ,  \label{eq:Vp(t)} \\
   \nonumber \\
  a_P \,\, &=&  \,\, c \, H_0 . \label{eq:Ap}
 \end{eqnarray}
 These would represent the anomalies found in the analysis
 of the spacecraft tracking data. However, with the value\cite{Anderson2002} of
 $a_P = (8.74 \, \pm \, 1.33) \times 10^{-8} \,{\rm cm} / {\rm s}^2$ this would require
 that
 $H_0 = a_P / c = (90.0 \, \pm \, 13.7) \,\, {\rm km} / {\rm s} / {\rm Mpc}$,
 measured at a redshift $z = 0$.  This is $6\, \%$ to  $44\, \%$ larger than the
 currently accepted value of $72 \,\, {\rm km} / {\rm s} / {\rm Mpc}$.  (It is not unlikely
 that if the spin histories (Ref. \cite{Anderson2002}, Sect. D) of both spacecraft were
 properly accounted for\cite{Mbelek-1}, then the acceleration related to Hubble's constant
 would be closer to $a_{P}(0) = (7.84 \pm 0.01) \times 10^{-8} \,{\rm cm} / {\rm s}^2$
 which would correspond to $H_0 = (80.1 \pm 0.1) \, {\rm km} / {\rm s} / {\rm Mpc}$.)
 The relation (\ref{eq:Ap}) for $a_P$ corresponds in principle to a minimum acceleration
 in the universe\cite{Carmeli2006} having a finite value $\, a_{\rm min} = c\, / \,\tau$,
 where the Hubble-Carmeli time constant $\tau \approx H^{-1}_0$.

% That said, we can look at the data plots.
% Ref. \cite{Anderson2002}, Fig. 8 shows the anomaly in the Pioneer 10 velocity, which agrees
% with (\ref{eq:Vp(t)}) if $H_0 = 90 \,\, {\rm km} / {\rm s} / {\rm Mpc}$.  Similarly,
%  Ref. \cite{Anderson2002}, Fig. 7 shows the anomalous acceleration of both spacecraft,
% for which (\ref{eq:Ap}) is a good representation for $20 \, {\rm AU}$ and beyond.
% For distances between $5$ to $20 \, {\rm AU}$ there is not good correlation, but
% these are distances where the solar radiation pressure
% may be interfering in the analysis (see Ref. \cite{Anderson2002}, Fig. 6).

\section{Transmission of light beam between a body bound to the solar system and an unbound body}

 Let us review the Pioneer 10 \& 11 anomaly with a little more detail.
 From our postulate (P-1) and the extended Hubble law (\ref{eq:massive_particle_Law}), we
 analyse the frequency changes in a light (radio) signal sent between the solar system barycenter
 (SSB) and a spacecraft which is in unbound motion near to the solar system. Since we hypothesize
 (H-1) that the solar system has escaped the Galaxy, then the unbound spacecraft is also
 galactically unbound. Thus, the spacecraft motion relative to the solar system is described
 by (\ref{eq:pioneer_eq_2}), though more precisely by general relativity theory plus the extra term
 due to the expansion.  The bodies in the solar system are bound and obey Newton's equation
 of motion (\ref{eq:Newtons_law}), though more precisely general relativity theory.
 
 The frequency shift upon the photons in the navigation beam due to gravity cancels out during
 a round trip. However, because of their speed $c$, these photons are unbound even to the
 Galaxy so that their frequencies will decrease according to (\ref{eq:photon_Law}).
 This cosmological redshift in the beam frequency in open space is actually independent
 of the H-1 hypothesis.

 For simplicity of argument, we assume the unbound spacecraft is moving radially outward
 from the SSB at a velocity $v(t)$ based on the parametrized post-Newtonian approximation (PPN)
 and any other known effects upon objects in deep solar system orbit, such as due from
 solar radiation pressure and planetary dust. With respect to the observer fixed to the
 SSB the entire round trip of a light beam of initial frequency $\nu_0$ originating
 at the SSB, traversing the distance to the spacecraft, reflecting off of the
 spacecraft and traveling back and received at the SSB, will have its frequency
 transformed on the outward journey by the factor $( 1 - z_1)$ due to the effect of the expansion
 on the photon beam signal, where $z_1 = H_0 \, \Delta{t_1}$ is the redshift during the travel
 time $ \Delta{t_1}$ to the spacecraft. At the spacecraft  the beam is received and retransmitted,
 incurring the double factor $( 1 - v(t)/c - v_{z}(t)/c)^2 $ in change to its frequency, where from
 (\ref{eq:v_z}), $v_{z}(t)= - c \, H_0\,t$ is directed parallel to the spacecraft time dependent
 velocity $v(t)$. On the return trip the beam is again altered by the expansion factor of
 $( 1 - z_2 )$, where $z_2 =  H_0 \, \Delta{t_2}$ is the redshift during the travel time
 $ \Delta{t_2}$ back to the SSB. The roundtrip effect is
\begin{eqnarray}
  \nu_{obs}(t) \,\, &=& \,\, \nu_0 \,  \left( 1 - z_1 \right) \left( 1 - v(t)/c - v_{z}(t)/c \right)^2
                \left( 1 - z_2 \right), \label{eq:nu_obs_begin} \\
  \nonumber \\
            \,\, &\approx& \,\, \nu_0 \, \left( 1 - 2 \, v(t) / c + 2 \,H_0 \, t
             - H_0 \Delta{t_1}  - H_0 \Delta{t_2}  \right) , \label{eq:nu_obs_final}
\end{eqnarray}
to first order in $v(t)/c$, $v_{z}(t)/c$, $z_1$  and $z_2$. If we limit the transmission times
to and from the spacecraft to half day each of $\Delta{t} \approx 5 \times 10^4 \,{\rm s}$,
with $H_0 = 90 \, {\rm km} / {\rm s} / {\rm Mpc}$
 $ = 2.916 \times 10^{-18}\,{\rm s}^{-1}$,
 then $z_1 \approx z_2 \approx H_0 \, 5 \times 10^4 \approx 1.46 \times 10^{-13}$.
Whereas, for $t$ a duration of a year or more, $v_{z}(t)/c = H_0 \, t > 9.2 \times 10^{-11}$.
Thus for this analysis we can ignore the effects of $z_1$ and $z_2$ on the beam signal
 since they are more than two orders of magnitude smaller than $v_{z}(t)/c$.

If we define the model expected observed frequency of the received light beam
\begin{equation}
  \nu_{model}(t) \,\, = \,\, \nu_0 \, \left( 1 - 2 \,v(t) / c \right)  , \label{eq:nu_model}
\end{equation}
then with this and (\ref{eq:nu_obs_final}) we have, in the Deep Space Network (DSN)
 negative format (Ref. \cite{Anderson2002}, Eq. 15), the anomalous frequency (blue) shift
\begin{equation}
  \Delta{\nu(t)}_{DSN} \,\, = \,\, -\left( \nu_{obs}(t) -  \nu_{model}(t) \right)
                    \,\, = \,\, - 2\, \nu_0  \,H_0 \, t ,
    \label{eq:nu_obs-nu_model} 
\end{equation}
which agrees with the Pioneer 10 \& 11 result (Ref. \cite{Anderson2002}, Eq. 15) if we have that
\begin{equation}
   H_0 \,\, = \,\, a_P / c . \label{eq:H_0=a_P/c}
\end{equation}

\section{Transmission of light beam between two bound solar system bodies}

 For bodies bound in the solar system the separation between the bodies are determined
 by general relativity, there being no further separation of the bodies nor perturbation
 of the orbital periods due to the expansion of the universe.
 The PPN determined velocity $v(t)$ for this case is only the radial component along the
 line of sight from the SSB to the bound spacecraft, as this is the component which
 contributes in the Doppler effect upon the beam. As before, the photons in the navigation
 beam are not bound to the Galaxy and so are subject to the effect of the expansion.

 Following the same line of reasoning as before, the round trip effect on the beam
 frequency is, from (\ref{eq:nu_obs_begin}) with $v_z(t) = 0$,
 \begin{eqnarray}
  \nu_{obs}(t) \,\, &=& \,\, \nu_0 \,  \left( 1 - z_1 \right) \left( 1 - v(t)/c \right)^2
                \left( 1 - z_2 \right),  \label{eq:nu_obs_bound_begin} \\
  \nonumber \\
            \,\, &\approx&  \,\, \nu_0 \, \left( 1 - 2 \, v(t) / c
                  - H_0 \Delta{t_1} - H_0 \Delta{t_2} \right) ,
              \label{eq:nu_obs_bound_final}
\end{eqnarray}
to first order in $v(t)/c$, $z_1$ and $z_2$.  Aside from the effects of $z_1$ and $z_2$,
this is the usual result obtained between bounded bodies in the solar system.
From (\ref{eq:nu_obs_bound_final}), the anomalous frequency (red) shift is (in DSN format)
\begin{equation}
  \Delta{\nu(t)}_{DSN} \,\, = \,\, -\left[ \nu_{obs}(t)
               - \nu_0 \, \left( 1 - 2 \, v(t) / c \right) \right]
              \,\, = \,\, \nu_0 \, H_0 \Delta{t}, \label{eq:nu_anom}
\end{equation}
where $\Delta{t} =\Delta{t_1}  + \Delta{t_2}$ is the round trip light travel time.
The current accuracy of atomic clocks\cite{NIST-F1} is better than one part in $10^{15}$.
If a spacecraft is placed in a circular orbit of 28 AU about the Sun (within Neptune's orbit,
where the solar radiation pressure acceleration will be $< 5 \times 10^{-8} {\rm cm} / {\rm s}^2$
for Pioneer 10 type spacecraft/antenna) the round trip travel time of a light signal sent to the
spacecraft would be about $\Delta{t} \approx 2.8 \times 10^4 \,{\rm s}$. From (\ref{eq:nu_anom})
 this corresponds to an anomalous frequency redshift ratio of the returned signal of
\begin{equation}
   \Delta{\nu(t)}_{DSN} / \nu_0  \, = \, H_0 \,\Delta{t}
   \, \approx \, \left( 2.916 \times 10^{-18}\,{\rm s}^{-1} \right) \, 2.8 \times 10^4 \,{\rm s}
      \, \approx \, 8.2 \times 10^{-14} , \label{eq:Delta_nu_by_nu0}
\end{equation}
which is 82 times larger than the clock accuracy. This would be detectable on a
 statistical basis. 

\section{Conclusion}

 It seems natural to extend the Hubble law to the realm of massive particles by way
 of the associated de Broglie wave of the particle. Then unbound particles would
 exhibit shifts in their wavelengths due to the expansion of the universe. This shift
 would be detectable as a decrease in momentum (velocity) of the particle. It is a
 claim of this paper that the anomalies found in the Pioneer spacecraft navigation is
 a detection of this cosmological velocity shift. But, in order for the spacecraft to
 be unbound galactically it was necessary to hypothesize (H-1) that the solar system
 is not bound to the Galaxy. This hypothesis is based on the fact that there appears
 to be an insufficient amount of visible mass in the Galaxy to bind the solar system.
 Thus the Pioneer spacecraft anomaly may be a counter example to the hypothesis that
 large amounts of dark matter exists in the Galaxy. A further effect, which is
 independent of the H-1 hypothesis, would be a cosmological redshift in the frequency
 of the communication signal in open space.

\end{document}